\documentclass[11pt,preprint]{aastex}
\usepackage{epsfig}
\begin{document}

\title{Deconfinement Phase Transition Heating and Thermal Evolution of Neutron Stars}
\author{Kang Miao$^{1,2}$, Pan Na-Na$^{2}$, Wang Xiaodong $^{1,2}$\\
1 - {\it The college of physics and electron,Henan university},
Kaifeng, 475004, Henan, P.R.China.\\
 2 - {\it  The institute of
astrophysics,Huazhong normal university},  Wuhan 430079, Hubei,
P.R.China.}


\begin{abstract}
 The deconfinement phase transition will
lead to the release of latent heat during spins down of neutron stars
if the transition is the
first-order one.
We have investigated the thermal evolution of neutron stars
undergoing such deconfinement phase transition. The results show
that neutron stars may be heated to higher temperature.This
feature could be particularly interesting for high temperature of
low-magnetic field millisecond pulsar at late stage.
\end{abstract}

\maketitle

Probing the equation of state (EOS) of neutron star matter is a urgent task in studying compact stars.
Both the determination of mass-radius relation\cite{07alf}\cite{04lat} and spin evolution of compact stars
\cite{00mad}\cite{03zhe}\cite{06pan}\cite{06zhe}\cite{05zhe}\cite{04zhe}are usual methods.
Neutron stars (NSs) cooling is another important tool for the study of
dense matter\cite{05pag}\cite{05liu}. By comparing cooling models with thermal emission
data from observations, we can gain insight into the EOS of dense matter inside NSs. Heating source inside the
stars is an important factor effect in the cooling of NSs.

It is known that a NS will spin down due to braking (e.g.
electric-magnetic radiation or gravitational wave radiation). The
deconfinement phase transition from hadronic matter to quark
matter can continuously occur in NSs during the spins down. Many
investigations were interesting in the phase transition which is
of the first-order type\cite{84pis}\cite{87gav}. Such
deconfinement processes induce continuous release of latent heat.
The generation of the energy increases internal energy of the
star. It will be called deconfinement heating (DH). DH have been
investigated in strange stars, where neutron drops at the bottom
of a crust drip on to the quark matter surface to be
instantaneously dissolved into quark
matter\cite{99yua}\cite{06yuw}.

Most of the approaches to deconfined matter in NSs use a standard
two-phase description of EOS where the hadron phase and the
quark phase are modelled separately and resulting EOS of the
mixed phase is obtained by imposing Gibbs conditions for phase
equilibrium with the constraint that baryon number as well as
electric charge of the system are conserved\cite{92gle}. Assumed the
deconfinement phase transition is a first order phase transition,
the deconfinement processes should produce latent heat.

 Combining with the equation of rotating structure based on
 Hartle's perturbation approach\cite{67har},
we get the total latent heat release unit time for a star in ref\cite{07kan}
\begin{equation}
H_{dec}(t)=\int \frac{de}{d\nu}\dot{\nu}(t)\rho_{B}dV.
\end{equation}
We can approximate equation (1)
 using a expression with a parameter $q_{n}$\cite{07kan1}
 \begin{equation}
H_{dec}(t)=q_{n}\frac{dN_{q}}{d\nu}\dot{\nu}(t)
\end{equation}
here $q_{n}$ is the average value of release energy per necleon
transforming into quarks, $N_{q}$ is deconfinement baryon number,$
\dot{\nu}$ is induced by magnetic dipole radiation. In a strange star with nuclear matter crust,
the similar expression has been obtained. However, $q_{n}$ in our model is $10^{-2}$ times smaller than
that of the strange star model\cite{07kan}\cite{01zdu}.

The cooling is realized via two channels - by neutrino emission
from the entire star body and by transport of heat from the
internal layers to the surface resulting in the thermal emission
of photons. Neutrino emission is generated in numerous reactions
in the interiors of neutron stars, as reviewed, by Page et al.
\cite{05pag}. In this paper, we consider the most powerful
neutrino emission including nucleon direct Urca (NDU) processes
and quark direct Urca (QDU) processes for the matter of NSs.
Nucleon superfluidity and quark superconductivity are not included
in the model.

According to the thermal evolution equation\cite{05pag}, we get the thermal
evolution curves with DH shown in Fig.1  which
shows the cooling behavior of a 1.5 $M_{\odot}$ NS
 for different magnetic fields ($10^{8}-10^{12}$G), where $q_{n}$ is taken to be 0.1 MeV.
 It is evident that the DH increase the
 surface temperature dramatically. This is extremely different from
 fast cooling scenario (solid curve). The
 strong field strength induces a rapid spin-down at the beginning
 while the low field strength leads to only obvious spin-down at
 the older ages.
In the cases of weak field, stars could maintain high temperatures
even at older ages ($>10^{7}$yrs). The temperature is nearly
identified with the values observed from millisecond pulsars,
especially for PSR J0437-4715\cite{04kar}.

\begin{figure}
 \centering
 \includegraphics[height=80mm]{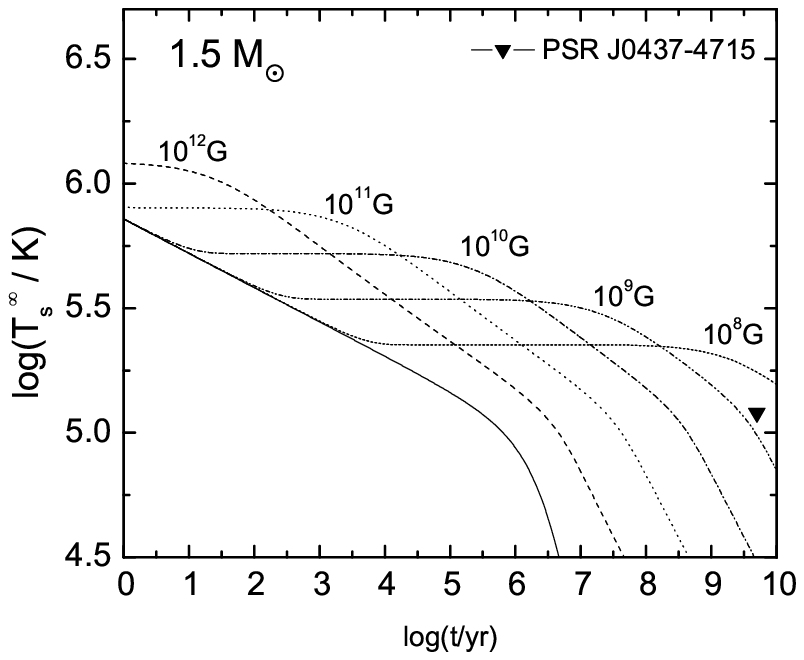}
 \caption{ Cooling curves of 1.5 $M_{\odot}$
neutron star with DH for various magnetic fields and the curves
without DH (solid curve).}
\end{figure}

 We find DH's
significant effects on the thermal evolution and the effects is
much more important than the past present heating
mechanisms\cite{95rei}\cite{92che}\cite{95van}\cite{06yuy}. We can find the remarkable change of temperature
 in strange stars nuclei where DH leads to too hot millisecond pulsars\cite{06yuw}\cite{06zho}.
In future, we
expect more observational examples in investigating effects of DH
mechanism on the NSs thermal evolution. Another problem which
remains to be investigated is the unified description of
middle-age and old pulsars. The NSs containing MP matter model may
be no bad selection when our combining DH,
superfluidity effects in nuclear matter together.

\begin{acknowledgements}
This work supported by NFSC under Grant Nos.10603002. I would like
to thank Prof X.P. Zheng for the useful disscussion.
\end{acknowledgements}

\end{document}